\documentstyle[epsf]{article}
\begin{document}
\title{Mixed state dense coding and its relation to entanglement measures}
\author{S. Bose $^1$, M. B. Plenio $^1$ and V. Vedral $^2$
\\ $^1$Optics Section, The Blackett Laboratory, Imperial College, \\ London SW7 2BZ, England \\
$^2$ Centre for Quantum Computing, Clarendon Laboratory,\\
	University of Oxford,
	Parks Road,
	Oxford OX1 3PU, England}

\maketitle
\begin{abstract}
Ideal dense coding protocols allow one to use prior maximal entanglement
to send two bits of classical information by the
physical transfer of a single encoded qubit. We investigate the case when
the prior entanglement is not maximal and the initial state of the 
entangled pair of qubits being used for the dense coding is a mixed state. We find upper and lower bounds on the capability to do dense coding in terms
of the various measures of entanglement. Our results can also be
reinterpreted as giving bounds on
purification procedures in terms of dense coding capacities.  
\end{abstract}

\section{Introduction}
One of the many suprising applications of shared entanglement \cite{mar} is  superdense coding introduced by Bennett and Wiesner \cite{wei}.
In the simplest example of this protocol, two people (Alice and Bob) share a pair of entangled qubits
(spin half particles or any other two state systems) in a Bell state.
Alice can then perform any of the four unitary operations given by
the identity ${\bf I}$ or the Pauli matrices ${\bf {\sigma_1}} ({\bf {\sigma_x}}), {\bf {\sigma_2}} ({\bf {\sigma_y}})$
and ${\bf {\sigma_3}} ({\bf {\sigma_z}})$ on her qubit. Each of these four unitary 
operations map the initial state of the two qubits to a different member
of the Bell state basis. Clearly, these four orthogonal and therefore fully distinguishable states can be used to encode two bits of information. After encoding her qubit, Alice sends it off to Bob who can extract these two bits of information
by performing a joint measurement on this qubit and his original qubit. This
apparent doubling of the information conveying capacity of Alice's
qubit because of prior entanglement with Bob's qubit is referred to as
superdense coding (which we shall henceforth refer to as just
dense coding). Dense coding has been implemented
experimentally with polarization entangled photons \cite{zei}. Some generalizations of the scheme to pairs of entangled N-level systems in
non-maximally entangled states (as opposed
to qubits in Bell states) \cite{Bar,haus} and to distributed multiparticle
entanglement \cite{bs} have also been studied.

     In the simplest example, the power of the method 
stems from the accessibility of all the four 
possible Bell states through local operations done by Alice alone on her
qubit \cite{wei,zei}. Obviously this accessibility is related
to the fact that the qubits shared by Alice and Bob are
in an {\em entangled} state. So an important question to ask is: {\em In what way does 
the capacity of conveying information through dense coding depend upon the degree of entanglement
of the initial shared pair of qubits}? Barenco and Ekert \cite{Bar} 
and Hausladen {\em et al.} \cite{haus} have investigated this
question when the initial shared entangled state is a pure state. Their
analysis shows that the amount of information conveyed by the dense coding procedure decreases
monotonically from its maximum value (two bits per qubit) with the decrease of the magnitude of the shared entanglement. It becomes one bit per qubit when the
entanglement  becomes zero.

    Recently, there has been a lot of work in quantifying the entanglement
of systems in mixed states \cite{ef,vlat1,vlat2,vlat3}. These measures
of entanglement have their physical interpretation in the process of
entanglement purification \cite{pf,Deu,ben2}. A natural question to ask is: {\em in
what way does the capacity to do dense coding depend on the degree of
entanglement as specified by these measures?} Answering such a question,
in effect, would mean {\em linking up the apparently disconnected concepts of
purification and channel capacity}. In this paper we take
a step in this direction by obtaining
bounds on the number of bits of information conveyed per qubit (let us call this {\bf C} after capacity) during dense coding in terms of the various measures of entanglement. It
should be noted that {\bf C} is a classical capacity as it quantifies
number of bits of information, but the carriers of the information are 
quantum (qubits).

    We also invert our results so that the value of {\bf C}
gives us information about the range within which the initial shared
entanglement lies. In other words, bounds on entanglement measures
(and hence on purification procedures) can be expressed in terms of
the classical capacity {\bf C}. This might be helpful, because, as
we shall show, the classical capacity for dense quantum coding is
readily calculable for certain special classes of dense coding protocols.


\section{The Holevo function as a measure of C}
  
    Suppose, Alice has a set of mixed
quantum states {$W_i$} at her disposal to convey some classical information (sequences of ones
and zeros) to Bob. Each state $W_i$ can be regarded as a separate letter.
Also suppose that Alice sends the state $W_i$ to Bob with an {\em a priori}
probability $p_i$. The ensemble that Alice uses to communicate with Bob is 
therefore given by
\begin{equation}
\label{avg}
W = \sum p_i W_i.
\end{equation} 
In the above case, the average number of bits of information that 
Alice can convey to Bob 
per transmission of a letter state is bounded from above by the Holevo function \cite{khol1}
\begin{equation}
\label{holv1}
H = S(W)- \sum p_i S(W_i),
\end{equation}
where $S(\rho)=-\mbox{Tr}~\rho \log \rho $ denotes the von Neumann entropy of
the state $\rho$ (Here, and throughout the rest of the paper, $\log$ stands for $\log_2$). Using a different notation, the Holevo function can
be rewritten as
\begin{equation}
\label{holv}
H = \sum p_i S(W_i||W),  
\end{equation}
where
$S(\sigma || \rho )$ stands for $\mbox{Tr} \{\sigma \log \sigma - \sigma \log \rho \}$ 
and is known as the quantum relative entropy between the states $\sigma$ and
$\rho$ \cite{don}.
It was recently shown that this bound can be achieved in the limit of
an infinite ensemble by appropriate
Block coding (grouping together and pruning long strings of letter states to
represent messages) at Alice's end and appropriate measurements at Bob's end
\cite{Holevo2}. 
  
 Now consider the case of dense coding. Alice and Bob initially
share an entangled pair of qubits in some state $W_0$, which may be mixed. Alice then performs local unitary operations on her qubit to put this shared
pair of qubits in either of the states $W_0, W_1,  W_2$ or $W_3$. In general,
Alice may use a completely arbitrary set of unitary operations to generate
these states:
\begin{equation}
\label{cg}
W_i={\bf U_i}\otimes {\bf I}~W_0~{\bf U_i}\otimes {\bf I}.
\end{equation}
In the above equation, ${\bf U_i}$  acts on Alice's qubit and
${\bf I}$ acts on Bob's qubit.
By sending  her encoded qubit to Bob, Alice is essentially communicating with Bob using the states $W_0, W_1,  W_2$ and $W_3$ as separate letters. The number of
bits she can communicate to Bob using this procedure is thus bounded by
the holevo function $H$ given in Eq.(\ref{holv}). Moreover, if
some block coding is done on a large enough collection of qubits in 
addition to the dense coding, then the number of bits of information
communicated is equal to the Holevo function. We will thus take

\begin{equation}
\label{see}
\mbox{\bf C}= H,
\end{equation}
assuming that any additional necessary block coding
will automatically be done to supplement the dense coding. Exactly the same
assumption has been used in Ref.\cite{haus} to calculate the capacity
for dense coding in the case of pure letter states. Eqs.(\ref{cg}) and 
(\ref{see}) define the most general version of dense coding
and we shall refer to this as completely general dense
coding (CGCD).

  A simpler example of
dense coding is the case when the letter states are generated from the
initial shared state $W_0$ by

\begin{eqnarray}
\label{w0}
W_0={\bf I}\otimes {\bf I}~W_0~{\bf I}\otimes {\bf I},\\
W_1={\bf \sigma_1}\otimes {\bf I}~W_0~{\bf \sigma_1}\otimes {\bf I}, \\
W_2={\bf \sigma_2}\otimes {\bf I}~W_0~{\bf \sigma_2}\otimes {\bf I}, \\
W_3={\bf \sigma_3}\otimes {\bf I}~W_0~{\bf \sigma_3}\otimes {\bf I}.
\label{w1}
\end{eqnarray}
In the above set of equations, the first operator of the combination
${\bf \sigma_i} \otimes {\bf I}$ acts on Alice's qubit and the second operator
acts on Bob's qubit. We shall refer to this case (i.e when the letter
states are generated by Eqs.(\ref{w0})-(\ref{w1})) as simply 
general dense coding (GDC). The generality present in GDC is that Alice
is allowed to prepare the different letter states with unequal probabilities.
In other words, one has to use Eqs.(\ref{avg})-(\ref{see}) to estimate
the capacity {\bf C}.
  
  In the more special case when Alice not only generates the four letter states
according to Eqs.(\ref{w0})-(\ref{w1})) but also 
with equal probability, the ensemble is given by
\begin{equation}
\label{avg1}
W = \frac{1}{4} \sum_{i=0}^3 W_i.
\end{equation}
and the capacity becomes
\begin{equation}
\label{c}
\mbox{\bf C} = \frac{1}{4} \sum_{i=0}^3 S(W_i||W).  
\end{equation} 
We shall call this simplest case special dense coding (SDC).
Among all the possible ways of doing GDC, SDC is the optimal way to 
communicate when $W_0$ is a pure state (as we shall show in the next
section) or a Bell diagonal state. However, we do not know the optimal way
to communicate when $W_0$ is a completely general state and CGCD is
allowed. For most of our paper, we shall
obtain bounds on the classical capacity {\bf C} for SDC only. But we shall 
point out those results which are valid
for GDC and CGDC as well. 
Though the main aim of this paper is to establish bounds on 
the capacity ${\bf C}$ when the
letters are mixed states, we shall begin with a calculation of ${\bf C}$ 
for pure letter states.

\section{C for SDC with pure letter states}
\label{purer}
Consider the initial shared pure state $W_0$ to be,
\begin{equation}
\label{p0}
|\psi_0\rangle  =   (a|00\rangle + b|11\rangle ).
\end{equation}
Then, according to Eqs.(\ref{w0})-(\ref{w1}), the other letter states
are given by
 
\begin{eqnarray}
|\psi_1\rangle  =  (a|10\rangle + b|01\rangle ), \\
|\psi_2\rangle  =   -i (a|10\rangle - b|01\rangle ), \\
|\psi_3\rangle  =   (a|00\rangle - b|11\rangle ),
\label{p3}
\end{eqnarray} 
from which we obtain $W_i = |\psi_i\rangle \langle \psi_i|$. As all $W_i$
are pure states we have
\begin{equation}
S(W_i)=0.
\end{equation}
Thus from Eqs.(\ref{holv1}) and (\ref{see}) we have
\begin{equation}
\label{sw}
\mbox{\bf C}=S(W).
\end{equation}
We will consider only the case of SDC. Thus the ensemble used
is obtained from Eq.(\ref{avg1}) to be
\begin{eqnarray}
W &=& \frac{|a|^2}{2}|00\rangle\langle 00| + \frac{|b|^2}{2}|01\rangle\langle 01| \nonumber \\ &+& \frac{|a|^2}{2}|10\rangle\langle 10| + \frac{|b|^2}{2}|11\rangle\langle 11|.
\end{eqnarray}
Thus from Eq.(\ref{sw}) for the capacity {\bf C}, we get
\begin{eqnarray}
\mbox{\bf C} &=& - (|a|^2 \log \frac{|a|^2}{2}+ |b|^2 \log \frac{|b|^2}{2}) \\
&=& 1 - (|a|^2 \log |a|^2+ |b|^2 \log |b|^2).
\end{eqnarray}
Now we should recall that a good measure of entaglement for a pure
state of a system composed of two subsystems A and B is given by the
von Neumann entropy of the state of either of the subsystems \cite{ben2,pop}.
Let us call this measure the von Neumann entropy of entanglement and 
label it by $E_v$. Thus
\begin{equation}
E_v(|\psi\rangle \langle \psi|_{\mbox \scriptsize {A+B}})= S {\large (}\mbox{Tr}_{\mbox \scriptsize {A}}(|\psi\rangle \langle \psi|_{\mbox \scriptsize {A+B}}){\large )},
\end{equation}
where $\mbox{Tr}_{\mbox \scriptsize {A}}$ stands for partial trace over
states of system A.
Therefore, for all the states $W_i$, 
\begin{equation}
E_v(W_i) = - (|a|^2 \log |a|^2+ |b|^2 \log |b|^2).
\end{equation}
Thus,
\begin{equation}
\label{pure1}
\mbox{\bf C}=1+E_v(W_i).
\end{equation}
  
  We now prove that for pure
states, SDC (using all alphabet states with equal {\em a priori}
probability)
is the optimal way to communicate among all possible ways of
doing GDC (i.e when the letter states are generated 
by Eqs.(\ref{w0})-(\ref{w1})). Consider the general case when the states
$W_i$ are sent with probabilities $p_i$. Then from Eq.(\ref{avg}) we have
\begin{equation}
W=\rho_1+\rho_2,
\end{equation}
where

\begin{eqnarray}
\label{ro1}
\rho_1 &=& (p_0+p_3)|a|^2 |00\rangle\langle 00| + (p_0+p_3)|b|^2|11\rangle\langle 11| \nonumber \\ &+& (p_0-p_3)a ^{*}b|11\rangle\langle 00| + (p_0-p_3)a b^{*} |00\rangle\langle 11|,
\end{eqnarray} 
and
\begin{eqnarray}
\rho_2 &=& (p_1+p_2)|a|^2 |10\rangle\langle 10| + (p_1+p_2)|b|^2|01\rangle\langle 01| \nonumber \\ &+& (p_1-p_2)a ^{*}b |01\rangle\langle 10| + (p_1-p_2)a b^{*}|10\rangle\langle 01| .
\label{ro2}
\end{eqnarray}
Therefore, $\rho_1$ and $\rho_2$ form separate blocks inside the matrix
$W$ and
\begin{equation}
S(W)=S(\rho_1)+S(\rho_2).
\end{equation} 
Eq.(\ref{sw}) indicates that we need to choose the probabilities $p_i$ in such
a way that $S(W)$ is maximized. Density matrices with same diagonal elements have the highest von Neumann
entropy when the nondiagonal elements are zero. Applying this fact to
Eqs.(\ref{ro1}) and (\ref{ro2}) we get
 
\begin{eqnarray}
\label{nond1}
p_0=p_3, \\
p_1=p_2.
\label{nond2}
\end{eqnarray} 
Using Eqs.(\ref{nond1}) and (\ref{nond2}) in the expressions for $\rho_1$
and $\rho_2$ and calculating the entropy $S(W)$ gives
 
\begin{eqnarray}
S(W)&=&-\sum_{i=0}^1 2p_i(|a|^2 \log 2p_i|a|^2 + |b|^2 \log 2p_i|b|^2) \nonumber  \\ &=& -\sum_{i=0}^1 2p_i \log 2p_i - (|a|^2 \log |a|^2 + |b|^2 \log |b|^2), 
\end{eqnarray}
where the normalizations of the state amplitudes and the probabilities 
$p_i$ have been used. From analysis of von Neumann entropies it is well
known that the expression $-\sum_{i=0}^1 2p_i \log 2p_i$ has a maximum
value when both $p_0$ and $p_1$ are equal. Thus $S(W)$, and hence the
classical capacity {\bf C}
is maximized when
\begin{equation}
p_0=p_1=p_2=p_3=\frac{1}{4}.        
\end{equation}
Thus, among all the possible ways of performing GDC, SDC is the optimal way to communicate when pure states are being used
as letters. From the above result and Eq.(\ref{pure1}) we can conclude that in the case of GDC
with pure letter states we have
\begin{equation} 
\label{pure2}
\mbox{\bf C} \leq 1+E_v(W_i).
\end{equation}
This result (Eq.(\ref{pure2})) had also been obtained in Ref.\cite{haus} from logical
arguments. Following a procedure analogous to the above proof
it can be shown that for Bell 
diagonal letter states SDC is again the optimal way to communicate
among all the possible ways of doing GDC.

\section{Nature of the ensemble used in SDC}

 In order to derive a lower bound on the classical capacity
{\bf C} for SDC we need to prove a 
crucial lemma concerning the nature of the
ensemble used in SDC: 
  
   {\em For states $W_i$ defined
in accordance to Eqs.(\ref{w0})-(\ref{w1}), the ensemble $W$ for SDC as defined
in Eq.(\ref{avg1}) is a disentangled state irrespective of the nature
of $W_0$}.

   To prove this, we have to start by assuming $W_0$ to be
a most general state of two qubits. This is given
by \cite{hor1}

\begin{eqnarray}
\label{wo1}
W_0 &=& \frac{1}{4}[I\otimes I + \sum_m r_m \sigma_m \otimes I  \nonumber \\
&+& I\otimes \sum_m s_m \sigma_m + \sum_{m,n} t_{mn} \sigma_m \otimes \sigma_n],
\end{eqnarray}
where the indeces $m$ and $n$ take on values from $1$ to $3$.
Substituting $W_0$ from Eq.(\ref{wo1}) into Eqs.(\ref{w0})-(\ref{w1}) we 
get

\begin{eqnarray}
\label{w11}
W_1 &=& \frac{1}{4}[I\otimes I +r_1 \sigma_1 \otimes I- \sum_{m \neq 1} r_m \sigma_m \otimes I  
+I\otimes \sum_m s_m \sigma_m \nonumber \\ &+&\sum_{n} t_{1n} \sigma_1 \otimes \sigma_n - \sum_{m\neq 1,n} t_{mn} \sigma_m \otimes \sigma_n], \\ 
W_2 &=& \frac{1}{4}[I\otimes I +r_2 \sigma_2 \otimes I- \sum_{m \neq 2} r_m \sigma_m \otimes I   
+I\otimes \sum_m s_m \sigma_m \nonumber \\&+& \sum_{n} t_{2n} \sigma_2 \otimes \sigma_n -\sum_{m\neq 2,n} t_{mn} \sigma_m \otimes \sigma_n], \\
W_3 &=& \frac{1}{4}[I\otimes I +r_3 \sigma_3 \otimes I- \sum_{m \neq 3} r_m \sigma_m \otimes I  
+ I\otimes \sum_m s_m \sigma_m \nonumber \\&+&\sum_{n} t_{3n} \sigma_3 \otimes \sigma_n - \sum_{m\neq 3,n} t_{mn} \sigma_m \otimes \sigma_n]. 
\label{w12}
\end{eqnarray}
Using expressions for $W_i$ from Eqs.(\ref{wo1})-(\ref{w12}) in Eq.(\ref{avg1})
we get
\begin{equation}
W = I \otimes \frac{1}{4}[ I + \sum_m s_m \sigma_m].
\end{equation}
This is very clearly a disentangled state. A plausible physical argument to
support this result can be as follows. We know that an equal mixture of
the four Bell states is a disentangled state. So it seems highly likely
that an equal mixture of less entangled states
will be disentangled as well. This result is, of course,
valid only for SDC, where the equal probabilities of the letter states
result in the careful cancellation of all the entanglement carrying terms.
It allows us to easily derive a lower bound on {\bf C} for SDC in terms
of one of the measures of entanglement.

\section{The relative entropy of entanglement as a lower bound on C} 
        To investigate quantitatively the relationship between 
entanglement and {\bf C} for arbitrary mixed letter states, it is necessary to use some measure 
of entanglement for mixed states. One such measure of entanglement, the relative entropy
of entanglement, has been
introduced in Ref.\cite{vlat1}. It has been shown to have a statistical
interpretation \cite{vlat2} as well as a physical interpretation \cite{vlat3,rns}
in forming bounds on the process of entanglement purification \cite{ef,pf,Deu,ben2}.    We will use the symbol $E_R$ to represent this measure
of entanglement. For an
arbitrary  mixed state  $\sigma$ it is given
by 

\begin{equation}
E_R(\sigma) = \min_{\rho \in {\cal D}} S(\sigma||\rho),
\end{equation}
where ${\cal D}$ is the set of disentangled states. A property which has to
be
satisfied by any legitimate entanglement measure is its invariance under local unitary operations \cite{vlat3}.  $E_R(\sigma)$ is thus
invariant under local unitary operations of the state $\sigma$.

 As the state $W$ of the ensemble used for SDC 
has been shown to be a disentangled state $(i.e.~W \in {\cal D})$ in the previous section, 
we have

\begin{equation}
S(W_i||W)  \geq  \min_{\rho \in {\cal D}} S(W_i||\rho) = E_R(W_i).
\end{equation}

Using the above inequality in Eq.(\ref{c}) for {\bf C} of SDC we  get
\begin{equation}
\label{zeq}
{\bf \mbox{C}} = \frac{1}{4} \sum_{i=0}^3 S(W_i||W) \geq \frac{1}{4} \sum_{i=0}^3 E_R(W_i).  
\end{equation}
As each of the states $W_i$ are derived from $W_0$ via local unitary
operations only,
we have
\begin{equation}
\label{eq}
E_R(W_i) = E_R(W_0)
\end{equation}
for all values of $i$. Combining Eqs.(\ref{eq}) and (\ref{zeq}) we
get

\begin{equation}
{\bf \mbox{C}} \geq  E_R(W_0).  
\end{equation}
This means that the classical capacity {\bf C} for SDC is bounded from
below by the relative entropy of entanglement of the initial shared mixed
state. This result is however, not generalizable to GDC as it relies
crucially on $W$ for SDC being a disentangled state.  

\section{Average distinguishability as an upper bound on C}
In this section, we are going to
divert a little bit from the main theme linking {\bf C} to entanglement measures and point out
another interesting bound on {\bf C} stemming from the mutual distinguishability of the letter states $W_i$.  It is known that 
the relative entropy $S(W_i || W_j)$ is a kind of statistical measure of the distinguishability  between the quantum states $W_i$ and $W_j$ \cite{vlat2}.
Thus for GDC one can define an average mutual distinguishability function as

\begin{equation}
\delta(p_i,W_i) = \sum_i \sum_j p_i p_j S(W_i || W_j).
\end{equation}
We show below that this average mutual distinguishability function
forms an upper bound on {\bf C}. Putting $W$ from 
Eq.(\ref{avg}) into Eq.(\ref{see})  we obtain

\begin{eqnarray}
{\bf \mbox{C}} &=&  \sum_{i=0}^3 p_i~S(W_i|| \sum_{j=0}^3 p_j~W_j) \nonumber \\ &=&  \sum_{i=0}^3  p_i~S(\sum_{j=0}^3 p_j~ W_i||\sum_{j=0}^3 p_j~  W_j),
\label{c1}
\end{eqnarray}
where the factor $\sum_{j=0}^3 ~p_j$ has been inserted before $W_i$
in the second step as its value equals unity. We now use the joint
convexity property of the relative entropy 

\begin{equation}
	S( \sum \lambda_i \sigma_i ||\sum \lambda_i \rho_i) \le \sum \lambda_i S(\sigma_i || \rho_i)
	\label{convex}
\end{equation}
to expand the right hand side of Eq.(\ref{c1}) and obtain

\begin{equation}
{\bf \mbox{C}} \leq  \sum_{i,j=0}^3 p_i p_j~S(W_i || W_j) 
= \delta(p_i,W_i).  
\end{equation}
Note that the above result is completely general because neither
does it require the special class of local unitary operations given 
by Eqs.(\ref{w0})-(\ref{w1}), nor does it require the probabilities
to be uniform.
Thus the classical capacity for CGDC is bounded from above by
the average mutual distinguishability function. Note that we are
pointing  this out just as an interesting bound and it is not
linked to the main theme of the paper (relating entanglement measures
stemming from purification procedures to dense coding capacities).

\section{An upper bound on C in terms of the entanglement of formation}

In this section we find an upper bound on {\bf C} in terms of yet
another measure of entanglement, namely, the entanglement of formation
$E_F$. Consider a decomposition of an arbitrary mixed state 
$\sigma$ in terms of pure states $\sigma_i$ :
\begin{equation}
\label{dec}
\sigma =  \sum_i \lambda_i \sigma_i.
\end{equation}   
Then the entanglement of formation of this state is defined
as \cite{ef,wot}
\begin{equation}
E_F(\sigma) = \min \sum_i \lambda_i E_v (\sigma_i), 
\end{equation}
where the minimum is taken over all decompositions of $\sigma$ of the 
type given by Eq.(\ref{dec}). 
     
  The initial shared state $W_0$ used in dense coding, will, in general, have several decompositions
in terms of pure states. Let the particular decomposition from which its
entanglement of formation $E_F(W_0)$ is
calculated (referred to as the entanglement minimizing 
decomposition \cite{wot}) be 
\begin{equation}
\label{decw}
W_0 =  \sum_m q_m W_{0m},
\end{equation} 
where $W_{0m}$ are pure states and $q_m$ are probabilities.
As Eq.(\ref{decw}) gives the entanglement minimizing decomposition,
we have
\begin{equation}
E_F(W_0)=\sum_m q_m E_v(W_{0m}),
\end{equation}
while the normalization of the probabilities imply
\begin{equation}
\label{norm}
\sum_m q_m = 1.
\end{equation}

 As each of the signal states $W_i$ are derived from $W_0$ by local
unitary operations, they can be decomposed as 
\begin{equation}
\label{decwi}
W_i = \sum_m q_m W_{im}, 
\end{equation}
where each pure state $W_{im}$ is connected to the pure state $W_{0m}$
by exactly the same local
unitary operation as that which connects $W_i$ to $W_0$. As any legitimate
measure of entanglement has to remain invariant under local unitary
operations, we have
\begin{equation}
\label{im}
E_v(W_{im})=E_v(W_{0m}),
\end{equation}
and
\begin{equation}
\label{i}
E_F(W_i) = E_F(W_0).
\end{equation}
>From Eqs.(\ref{im}) and (\ref{i}) we have
\begin{eqnarray}
\label{equa}
\sum_m q_m  E_v(W_{im}) &=& \sum_m q_m E_v(W_{0m}) \nonumber \\
&=& E_F(W_0)=E_F(W_i).
\end{eqnarray}

Now, in the case of SDC, the capacity (from Eq.(\ref{c})) is
\begin{eqnarray}
{\bf \mbox{C}} &=& \frac{1}{4} \sum_i  S(W_i||\frac{1}{4} \sum_j W_j) \nonumber \\
 &=& \frac{1}{4} \sum_i S(\sum_m q_{m} W_{im}||\frac{1}{4} \sum_j \sum_m q_{m} W_{jm}) \\ &=& \frac{1}{4} \sum_i S(\sum_m q_{m} W_{im}|| \sum_m q_{m} W_m)
\label{grhl}
\end{eqnarray}
where we have put
\begin{equation}
\label{ensm}
W_m = \frac{1}{4} \sum_j W_{jm}.
\end{equation}
Using joint convexity (Eq.(\ref{convex})) to expand the right hand side
of Eq.(\ref{grhl}), we get
\begin{eqnarray}
\label{less}
{\bf \mbox{C}} & \leq & \frac{1}{4} \sum_i \sum_m q_{m} S(W_{im}|| W_m) \nonumber \\
&=& \sum_m q_{m} ( \frac{1}{4} \sum_i  S(W_{im}|| W_m) ). 
\end{eqnarray}
Now, for each value of $m$, the expression $\frac{1}{4} \sum_i  
S(W_{im}|| W_m)$ can be regarded as the classical capacity {\bf C} for SDC with the states
$W_{im}$ as the four letter states. As each of the
states $W_{im}$, are pure, Eq.(\ref{pure1}) implies
\begin{equation}
\label{oneE}
\frac{1}{4} \sum_i  
S(W_{im}|| W_m) = 1+E_v(W_{im}).
\end{equation}
Putting Eq.(\ref{oneE}) into Eq.(\ref{less}) we get
\begin{equation}
\label{qm1}
{\bf \mbox{C}} \leq \sum_m q_{m} (1+E_v(W_{im})).
\end{equation}
Using Eqs.(\ref{norm}) and (\ref{equa}) to simplify the right hand side
of Eq.(\ref{qm1}) we get
\begin{equation}
\label{upbnd}
{\bf \mbox{C}} \leq 1 + E_F(W_i).
\end{equation}
   
    The above bound is valid for GDC as well. To see this
one just has to repeat the above proof starting with 
\begin{equation}
{\bf \mbox{C}} = \sum_i p_i~S(W_i|| \sum_j p_j W_j)
\end{equation}
and replace Eq.(\ref{ensm}) with
\begin{equation}
\label{ensm1}
W_m =  \sum_j p_j~W_{jm}.
\end{equation}
Eq.(\ref{less}) then gets replaced by
\begin{equation}
\label{ii0}
{\bf \mbox{C}} \leq \sum_m q_{m} (\sum_i p_i~S(W_{im}|| W_m)).
\end{equation}
Now note the fact that $\sum_i p_i~S(W_{im}|| W_m)$ is the
expression for the classical capacity {\bf C} for CGDC with $W_{im}$
being the letter states. When the
states $W_{im}$ are generated from $W_{i0}$ according to
Eqs.(\ref{w0})-(\ref{w1}) (i.e. when GDC protocol is being followed),
then the purity of the states $W_{im}$ guarantees that
{\bf C} is less than $1+E_v(W_{im})$ (as shown
in section \ref{purer}). Using this fact in Eq.(\ref{ii0}) we again end up with Eq.(\ref{upbnd}). Thus even in the case of GDC, the capacity {\bf C}
is bounded by $1 + E_F(W_i)$.

\section{An upper bound on C in terms of the relative entropy of
entanglement}
Having analytically proven that $1 + E_F(W_i)$ is an upper bound
on {\bf C} for GDC, we now proceed to check whether the even
smaller (as proved in Ref.\cite{vlat3}) quantity $1+E_R(W_i)$ is
also an upper bound. However, we do not attempt to prove this analytically
for a completely general initial shared state $W_0$. Instead we calculate
the capacity {\bf C} (for SDC only) for those
specific classes of the initial shared state $W_0$ whose relative
entropy of entanglement $E_R(W_0)$ is already known.
We then plot
this capacity {\bf C} as a function of the relative entropy of entanglement 
$E_R$ for each of these classes of states and check whether
the curve ${\bf \mbox{C}}(E_R)$ lies below the plot of $1+E_R$. 

\begin{figure}
\begin{center} 
\leavevmode 
\epsfxsize=8cm 
\epsfbox{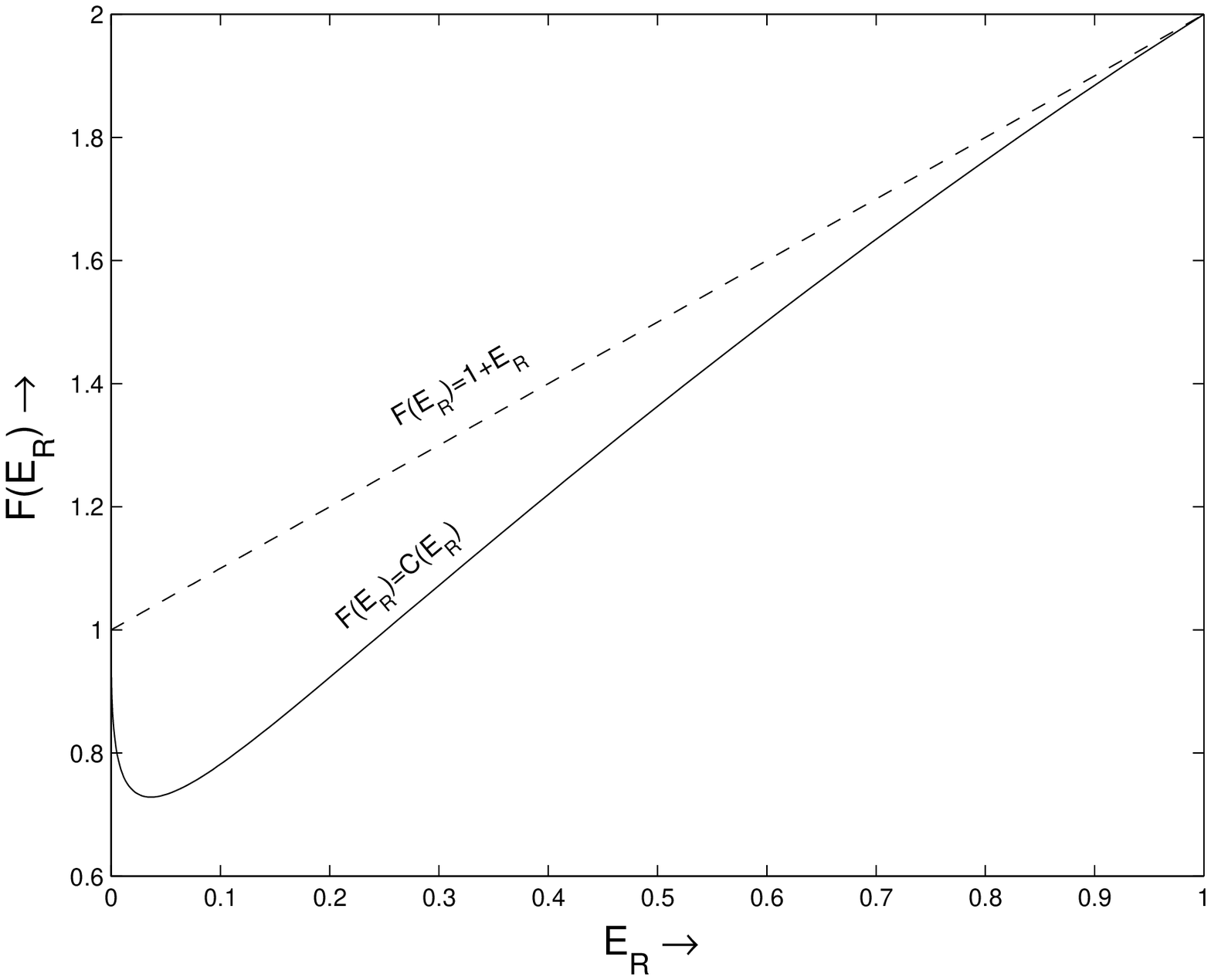}
\caption{Two different functions $F(E_R)$ of the
relative entropy of entanglement $E_R$ are plotted. The continuous line shows  the capacity {\bf C}
for SDC as a function of the relative entropy of entanglement $E_R$ if the initial
shared state is a lambda state of the type A. The dashed line shows the
function $1+E_R$. The figure illustrates that the capacity {\bf C}
for SDC is bounded
from above by $1+E_R$}
\label{lam1} 
\end{center}
\end{figure}

   At first we have a look at 
mixed states of the type \cite{vlat3}

\begin{equation}
\label{lamb}
\Lambda_A =  \lambda |\Phi^+\rangle \langle \Phi^+| +
(1-\lambda) |01\rangle\langle 01|,
\end{equation}
where $|\Phi^{+}\rangle$ is one of the four Bell states which are defined by
\begin{eqnarray}
|\Phi^{\pm}\rangle & = & \frac{1}{\sqrt{2}} (|00\rangle \pm |11\rangle ) \\
|\Psi^{\pm}\rangle & = & \frac{1}{\sqrt{2}} (|01\rangle \pm |10\rangle )
\end{eqnarray}
We shall refer to the states described by Eq.(\ref{lamb}) as 
lambda states of
the type A. For them, the relative entropy of entanglement is \cite{vlat3}
\begin{equation} 
 E_R(\Lambda_A)  =  (\lambda -2)\log (1-\frac{\lambda}{2}) + (1-\lambda)
\log (1-\lambda),  
\end{equation}
while the {\bf C} for SDC is

\begin{eqnarray}
{\bf \mbox{C}} &=&  (1-\lambda)
\log (1-\lambda) + \frac{1}{2}(\lambda -2)\log (1-\frac{\lambda}{2}) \\
&+& \frac{1}{2} \lambda \log \lambda + (1+ \frac{\lambda}{2}).
\end{eqnarray}
The plot of the classical capacity {\bf C} for these states as a function of $E_R$ of the state
has been shown in figure \ref{lam1} and it is indeed found that

\begin{equation}
{\bf \mbox{C}} \leq 1+E_R.
\end{equation}
The equality holds true only at the two ends of the graph, namely
at maximal entanglement $E_R=1$ (when $\sigma_1$ approaches
a Bell state) and zero entanglement $E_R=0$ (when $\sigma_1$ approaches
$|01\rangle\langle 01|$).

Next we look at states of the type \cite{vlat3}
\begin{equation}
 \Lambda_B  =  \lambda |\Phi^+\rangle \langle \Phi^+| +(1-\lambda) |00
\rangle\langle 00| 
\end{equation}
which we call lambda states of the type B. For these \cite{vlat3}
\begin{eqnarray}
\label{b1}
E_R(\Lambda_B)  &=&  s_+ \log s_+ + s_- \log s_-  \nonumber \\
&-& (1-\frac{\lambda}{2}) \log (1-\frac{\lambda}{2}) - \frac{\lambda}{2} \log \frac{\lambda}{2} \;\; ,
\end{eqnarray} 
where 
\begin{equation}
s_{\pm} = \frac{1\pm \sqrt{1-2\lambda (1-\lambda)}}{2}
\end{equation}
are the eigenvalues of $\Lambda_B$. The {\bf C} for SDC in this case
is given by                        
\begin{eqnarray}
\label{b2}
{\bf \mbox{C}}&=& s_+ \log s_+ + s_- \log s_-\nonumber \\
&-& (1-\frac{\lambda}{2}) \log \frac{1}{2}(1-\frac{\lambda}{2}) - 
\frac{\lambda}{2} \log \frac{\lambda}{4}.
\end{eqnarray}
As is clear from a simple comparison of
Eqs.(\ref{b1}) and (\ref{b2}), for lambda states of the type B,
\begin{equation}
\label{eq1}
{\bf \mbox{C}} = 1+E_R.
\end{equation}
Thus, a curious feature of lambda states of type
B is that the capacity for SDC is actually always {\em equal} to
$1+E_R$. We will discuss a bit more about this curious aspect later in
the section.
  
\begin{figure}
\begin{center} 
\leavevmode 
\epsfxsize=8cm 
\epsfbox{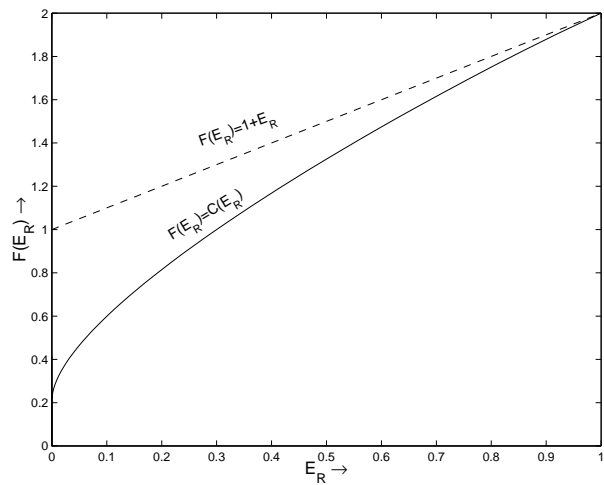}
\caption{The continuous line shows how the capacity {\bf C}
for SDC varies with the relative entropy of entanglement $E_R$ of the initial
shared Werner state. The dashed line shows that it is bounded
from above by $1+E_R$}
\label{wer} 
\end{center}
\end{figure}

    Now we consider another special class of states called
the Werner states \cite{ef,wer} parameterized by a number F called the 
fidelity \cite{ef} and given by  
\begin{eqnarray}
W_F&=&F|\Psi^{-}\rangle \langle \Psi^{-}| + \frac{1-F}{3} (|\Psi^{+}\rangle \langle \Psi^{+}| \nonumber \\ &+& |\Phi^{+}\rangle \langle \Phi^{+}|
+ |\Phi^{-}\rangle \langle \Phi^{-}|). 
\end{eqnarray}
The relative entropy of entanglement of these states is given by \cite{vlat3}
\begin{equation}
 E_R(W_F) = F\log F +(1-F)\log (1-F) + 1, 
\end{equation}
and the {\bf C} for SDC is calculated to be
\begin{eqnarray}
{\bf \mbox{C}} &=& 2 + F\log F +(1-F)\log \frac{(1-F)}{3}. 
\end{eqnarray}
The plot {\bf C} versus $E_R$ has been drawn in Fig.\ref{wer} and it is 
found that even in this case, 
\begin{equation}
{\bf \mbox{C}} \leq 1+E_R.
\end{equation}

Now consider the case of a general Bell diagonal state 
\begin{eqnarray}
B_D &=& \lambda_1 |\Psi^{-}\rangle \langle \Psi^{-}| + \lambda_2 |\Psi^{+}\rangle \langle \Psi^{+}| \nonumber \\
&+& \lambda_3 |\Phi^{+}\rangle \langle \Phi^{+}|
+\lambda_4 |\Phi^{-}\rangle \langle \Phi^{-}|.
\end{eqnarray}
It has been proved in Ref.\cite{vlat1} that when all $\lambda_i \in [0,\frac{1}{2}]$ then
\begin{equation}
\label{zero}
E_R(B_D)=0,
\end{equation}
while when any of the $\lambda_i$ (say $\lambda_1$) $\geq \frac{1}{2}$,
then
\begin{equation}
\label{fint}
E_R(B_D)= \lambda_1\log\lambda_1 +(1-\lambda_1)\log (1-\lambda_1) + 1.
\end{equation}
The {\bf C} for SDC always turns out to be
\begin{equation}
\label{ccc}
{\bf \mbox{C}}= 2 + \sum_i \lambda_i \log\lambda_i.
\end{equation}
First consider the case when all $\lambda_i \in [0,\frac{1}{2}]$. From
Eqs.(\ref{zero}) and (\ref{ccc}) we find that
\begin{eqnarray}
1+E_R(B_D)-{\bf \mbox{C}} &=& - \sum_i \lambda_i \log\lambda_i -1 \nonumber
\\ &=&  - \sum_i \lambda_i (\log\lambda_i+1) \nonumber \\
&\geq & 0,
\end{eqnarray}
where $\sum_i \lambda_i =1$ has been used to proceed from the first to
the second step and $ \log\lambda_i \leq -1$ (because $\lambda_i \in [0,\frac{1}{2}]$) has 
been used to proceed from the
second to the third step.

Now consider the complementary case ($\lambda_1 \geq \frac{1}{2}$). From
Eqs.(\ref{fint}) and (\ref{ccc}) we find that

\begin{eqnarray}
1+E_R(B_D)-{\bf \mbox{C}} &=& (1-\lambda_1)\log (1-\lambda_1) -\sum_{i\neq 1} \lambda_i \log\lambda_i \nonumber \\ &=& (\sum_{i\neq 1} \lambda_i) \log (\sum_{j\neq 1} \lambda_j)-\sum_{i\neq 1} \lambda_i \log\lambda_i \nonumber \\
&=& \sum_{i\neq 1} \lambda_i (\log (\sum_{j\neq 1} \lambda_j)- \log\lambda_i)
\nonumber \\ &\geq& 0,
\end{eqnarray}
where in order to proceed from the third to the fourth step we have used
the simple fact that $\log(\lambda_2+\lambda_3+\lambda_4)$ is greater than
either of the terms $\log \lambda_2$,$\log \lambda_3 $ or $\log \lambda_4$.
Thus for all Bell diagonal states we have
\begin{equation}
{\bf \mbox{C}} \leq 1+E_R.
\end{equation}
 
 Now we will point out a curious fact about the situation when any two of the eigenvalues of
a Bell diagonal state (say $\lambda_3$ and $\lambda_4$) are zero. When 
$\lambda_1=\lambda_2=\frac{1}{2}$ (which means $E_R=0$), we have {\bf C}$=1$
(using Eq.(\ref{ccc})).
In all the other cases (for which we use Eqs.(\ref{fint}) and (\ref{ccc})) we have

\begin{eqnarray}
1+E_R(B_D)-{\bf \mbox{C}} &=& (1-\lambda_1)\log (1-\lambda_1) - \lambda_2 \log
\lambda_2 \nonumber \\
&=& 0,
\end{eqnarray}
where we have used the simple fact that $\lambda_1+\lambda_2=1$. Thus for
all Bell diagonal states with only two nonzero eigenvalues we have
\begin{equation}
\label{eq2}
{\bf \mbox{C}} = 1+E_R.
\end{equation}

  On the basis of the results obtained in this section let us
 conjecture:

{\bf Conjecture }: {\em The {\bf \mbox{C}} for SDC with
completely general (possibly mixed) states is bounded from above by $1+E_R$}. 

        A great
deal of empirical evidence has been presented in this section in support
of the conjecture. In the next section we proceed to give a heuristic justification in support of our conjecture. In fact, we will try to
justify an even stronger upper bound on {\bf C}.

\section{SDC and purification}
\label{pff}
To justify the conjecture of
the previous section, we will have to examine the 
following interesting question:
How does the capacity {\bf C}
 change
if Alice and Bob first locally purify their ensemble and distill Bell states 
(following the optimal purification procedure) and
follow this up by SDC?  Various purification procedures
have been described in Refs.\cite{ef,pf,Deu,ben2}. Here we assume that
Alice and Bob follow the optimal purification process: One which helps
them to distill the maximum fraction of Bell states from the initial
ensemble.
They will, after optimal purification,
have a fraction $E_D$ (where $E_D$ is called the
{\em entanglement of distillation})
shared pairs in Bell states and a fraction $(1-E_D)$ pairs in
a disentangled state. They now complete their 'purification'
process by converting the final subensemble of disentangled pairs to pure
states by projective measurements. We refer to such a protocol
as {\em complete purification}. After a  complete purification,
Alice can use the fraction $E_D$ of Bell pairs to send Bob information
at the rate of 2 bits/pair and the fraction $(1-E_D)$ of
pure disentangled pairs to send information at
the rate of 1 bit/pair. Thus if Alice and Bob initially shared
$n$ pairs, the classical capacity after a complete
and optimal purification procedure is
\begin{eqnarray}
\label{cppd}
{\bf \mbox{C}}&=& \frac{1}{n} \{2 n E_D + n (1-E_D)\} \nonumber \\
&=& 1+E_D.
\end{eqnarray}
Note that the above result is only asymptotically true ($n \rightarrow \infty$).
Naively, one might expect this {\bf C} to be lower than the {\bf C}
before purification.
This is because, as mentioned in Ref.\cite{ben2}, entanglement
concentration is a more destructive process than quantum data
compression. Some amount of shared entanglement is destroyed in the process
of purification. So it might be expected that the final ensemble after a
purification will be able to convey less classical information
than the original ensemble. On the contrary, as we will justify, the
capacity for SDC with the purified ensemble is greater than that
with the unpurified ensemble when an optimal
and complete purification protocol is used. The gain comes from the fact that now
there are two separate sub-ensembles instead of a single ensemble.
 In this section, we will try to
justify, albeit heuristically, the following statement (as a more fundamental
conjecture than the one presented in the previous section):\\
{\bf Conjecture}: {\em The capacity for SDC is more when it is preceeded
by a complete and optimal purification procedure}.

 We will first consider the special case of Bell diagonal mixed states, for
which the proof of the increase in the SDC capacity on a 
complete and optimal purification (our conjecture) can be rigorously proved. For Bell diagonal states
with entropy $S(\rho) \leq 1$, there is a purification protocol
called hashing (with the distillable fraction being $E_{DH} = 1-S(\rho)$) \cite{ef}.
>From Eq.(\ref{ccc}) we see that for these Bell diagonal states, the {\bf C} for SDC
is equal to $1+E_{DH}$. As hashing may not necessarily be the optimal
protocol, we have $E_{DH} \leq E_D$. This immediately implies ${\bf \mbox{C}} \leq  1+E_D$. For the complementary case of Bell diagonal states with
$S(\rho) > 1$, we have (from Eq.(\ref{ccc}) for SDC), ${\bf \mbox{C}} < 1$.
This is obviously less than $1+E_D$ for any finite value of the entanglement
of distillation (i.e for all inseperable states \cite{hor2}). Thus for all
Bell diagonal states the {\bf C} for SDC can be improved by a prior optimal
and complete purification of the ensemble of shared states.

   For the more general case of arbitrary mixed entangled states, the 
proof will, essentially, be heuristic. Our approach will be to split the change in the classical information
capacity due to complete and optimal purification into two parts. The first
part is positive (an increase in capacity) and due to the addition of
classical side channels during the purification procedure. These side
channels are used by Alice (A) to communicate the results of her measurements
to Bob (B), or vice versa. As this communication during purification is
already conveying information from A to B (or vice versa), the channel
capacity of the classical side channels should be directly added to
the classical capacity (of course, for this, we have to implicitly assume
the additivity of classical capacities). The information conveyed from A to B (or vice versa)
during purification is actually used by the two parties to
precisely identify their entangled and disentangled subensembles. There is
also a negative contribution (decrease in capacity) during the purification
as a fraction $1-E_D$ of shared pairs loose all their entanglement. Of course
a part of this entanglement is pumped into the entangled subensemble, but the
remaining is lost. Our job will be to argue that the positive contribution
in channel capacity due to addition of classical side channels 
outweighs the negative
contribution due to loss of entanglement in an optimal and complete
purification procedure.

   Consider the
process of purification as a process of information gain 
 by A and B. Before the purification,
both A and B have an equal amount of knowledge about their
shared system (they both know the full density matrix). Finally, each of their shared states are pure (because of the completeness of the purification
procedure) and both of them have equal knowledge (i.e each know which of the pairs form
the disentangled subensemble and which of the pairs form the maximally entangled
subensemble). Thus they have both  
gained an equal amount of information $\mbox{I}$/pair about their shared pairs during
the process of purification. Now, acquiring a part of this information
may not cause any destruction of shared entanglement (lets call it $\mbox{I}_1$/pair),
while the remaining part (lets call it $\mbox{I}_2$/pair) does cause
a lowering of shared entanglement. We now contend, that this part, 
$\mbox{I}_2$/pair has to be equal
to the number of classical communication channels used in the optimal
purification procedure. The logic follows from the fact that the optimal
strategy would be for A and B to collaborate in such a way that both of
them gain
the entire information $\mbox{I}_2$/pair with the least destruction
of entanglement. They must then each acquire only a fraction of the
information complementary to the fraction acquired by the other. In this
way, the total information acquired by A and B from the shared pairs through
direct entanglement degrading measurements is $\mbox{I}_2$/pair. They can then use a minimum of $\mbox{I}_2$ classical side channels per pair to
communicate their fraction of the acquired information to the other party.
On the other hand, if each wanted to acquire a fraction of the total information
by direct measurements which was not entirely complementary to the part
acquired by the other, they would destroy more entanglement than is really
necessary. Thus we would expect that the classical side channels contribute to boosting the
capacity up by at least an amount $\mbox{I}_2$/pair in the optimal purification
procedure. 

    Now consider how much the capacity decreases due to the degradation of 
shared entanglement during the purification procedure. The classical capacity
of a fraction $1-E_D$ of the pairs drops from {\em at most} 2 bits/pair to
1 bit/pair (due to the {\em completeness} of the purification procedure, the
final capacity cannot be lower than 1 bit/pair). Thus the drop in
classical capacity of each of the finally disentangled pairs 
is not more than 1 bit. This implies that the net decrease in classical
capacity of the entire ensemble 
due to loss of entanglement hasn't been more than $1-E_D$ bits/pair.

    When the information $\mbox{I}_2$ is greater than 1 bit/pair, the proof of our
statement is straightforward. The increase in capacity due to classical
side channels (i.e $\mbox{I}_2$/pair) is more than 1 bit/pair, while the
decrease in capacity due to loss of entanglement is less than $1-E_D$ bits/pair.
As $1-E_D$ is a fraction (i.e $1-E_D < 1$), the increase in capacity on
purification overrides the inevitable decrease in capacity due to degradation
of shared entanglement. Clearly, thus the capacity to do SDC increases
on prior purification when $\mbox{I}_2 \geq 1$ bit.

     Now, we have to show that no matter what the value of $\mbox{I}_2$ is,
it is {\em possible} to find a purification protocol for which $\mbox{I}_2$ is greater than $1-E_D$ bits. In other words, if initially there
are $n$ shared pairs (with $n$ being large), then the
 number of
shared pairs which loose all their entanglement on gaining an
information of $\mbox{I}_2$/pair can be made less than $n \mbox{I}_2$ by
an appropriate choice of the purification protocol. From such a statement it directly
follows that the increase in capacity on
purification overrides the inevitable decrease in capacity due to degradation
of shared entanglement. A simple way in which one may try to
justify the above  proposition is from the fact that one
pair generally allows the extraction of upto two bits of information (one bit from each of A's 
and B's qubit). So, it should be possible, in general, to extract $n \mbox{I}_2$  amount of
information from measurements on less than $n \mbox{I}_2$ shared pairs (whose
entanglement is degraded).
This intutive understanding leaves
open the question as to whether the information $\mbox{I}_2$/pair {\em relevant} to purification
can be acquired in this way. For that we resort to explicit exemplification.
Consider the class of mixed states $\rho$ (having nonzero entropy  $S(\rho)$)
and a diagonal decomposition $\sum \lambda_i |\psi_i\rangle \langle \psi_i|$ in terms of the pure states $|\psi_i\rangle$. Also suppose that this
diagonal decomposition does not coincide with the entanglement minimizing
decomposition \cite{wot} of $\rho$. For such states, acquiring the information
about the entropy ($S(\rho)$/pair) is essential during
purification, as the final ensemble is pure. We now contend that acquiring this
$S(\rho)$ /pair of information {\em necessarily} destroys some
amount of shared entanglement. We will justify this by the method 
of contradiction. Suppose, it was really possible to acquire the information
about the entropy $S(\rho)$ without destroying any entanglement. On acquiring an information of the amount equal to $S(\rho)$/pair,  A and B
will be able to divide their initial ensemble to four seperate pure subensembles,
each being comprised of one of the states $|\psi_i\rangle$. The weight of each of these subensembles will be equal to
the eigenvalues $\lambda_i$. The average entanglement shared by A and B after extracting the information $S(\rho)$/pair is thus $\sum \lambda_i E_{v}(|\psi_i\rangle)$.  However, for the class of states $\rho$
that we are considering, $\sum \lambda_i E_{v}(|\psi_i\rangle)$ is necessarily greater than the initial
shared entanglement (as quantified by the entanglement of formation $E_F$).
Thus by purely local actions and classical communications,
A and B have been able to increase the shared entanglement for these
classes of mixed states. This contradicts
the definition of entanglement as a quantity that cannot be increased by 
local actions and classical communications. Thus, {\em mixed states with the diagonal
decomposition not coincident with the entanglement minimizing decomposition cannot be locally purified
 without necessarily destroying some shared entanglement}. Thus the information
about the entropy of these states is of the type $\mbox{I}_2$ (i.e acquiring the information necessarily causes
degradation of entanglement). A corollary which immediately follows (though
not directly relevant to the main point of the paper), is: {\em the optimal purification
of mixed states with the diagonal
decomposition not coincident with the entanglement minimizing decomposition necessarily has a nonzero $\mbox{I}_2$ and thus necessarily
requires the use of classical side channels}. Now consider such states when
their entropy, and thereby $\mbox{I}_2$, exceeds $1$ bit. Thus the total information $n \mbox{I}_2$ needed to
be acquired is greater than 
$n$ bits. However, purifiability (nonzero $E_D$) implies that the total entanglement loss
can be entirely concentrated to the entanglement loss of $n(1-E_D)$ pairs which is less than $n \mbox{I}_2$ pairs.
Therefore, it is {\em possible} to gain the relevant $n \mbox{I}_2$ bits of information by
destroying the entanglement of less than $n \mbox{I}_2$ pairs. Now we really
need an unproven extension of the above statement: {\em even when} $\mbox{I}_2
\leq 1$ bit and {\em irrespective of the nature of the decomposition of}
$\rho$, it is {\em possible} to gain the relevant $n \mbox{I}_2$ amount of information by
destroying the entanglement of less than $n \mbox{I}_2$ pairs. Though this
may seem a large extrapolation, it seems highly {\em plausible} because one
shared pair allows the extraction of upto two bits of information. Thus we would
expect that, no matter what the value of $\mbox{I}_2$ is, the increase in capacity on
purification overrides the inevitable decrease in capacity due to degradation
of shared entanglement.

    Thus the capacity to do SDC with the two  pure subenembles (maximally entangled and fully disentangled) after a complete and optimal purification, is
expected to be greater
than the initial mixed (and uniformly entangled) ensemble. We thus conjecture,
with the help of Eq.(\ref{cppd}), the
following bound on the capacity of
doing SDC with any mixed entangled state

\begin{equation}
\label{bnded}
{\bf \mbox{C}} \leq  1+E_D.
\end{equation}

This is an even stronger upper bound on the classical capacity than the bound
conjectured in the previous section as the relative entropy of entanglement
$E_R$ is necessarily greater than $E_D$ \cite{vlat3}. From Eq.(\ref{bnded}),
it directly follows that ${\bf \mbox{C}} \leq  1+E_R$ (our previous conjecture). 
Note that in this section, this conjecture has been rigourously proved for all Bell diagonal states,
heuristically proved for all states with entropy $S(\rho) \geq 1$ bit (i.e $\mbox{I}_2 \geq 1$ bit) and shown to be highly plausible for other types of
states.


\section{Conclusion}
In this paper we have obtained bounds on the capacity of doing
dense coding in terms of the different measures of entanglement stemming
from purification procedures. The
 rigorously proved part of the results of this paper is that for SDC one always has
\begin{equation}
 E_R \leq {\bf \mbox{C}} \leq 1+E_F.
\end{equation}
On the other hand if we allow for a conjecture (well supported by
examples and heuristically justified in Section.\ref{pff}), we have the stronger result
\begin{equation}
 E_R \leq {\bf \mbox{C}} \leq 1+E_R.
\end{equation}
What is the significance of bounds on the {\bf C} for SDC when it can be
readily calculated? The importance can be realized when we invert 
 the above equation and write
\begin{equation}
 {\bf \mbox{C}}-1 \leq E_R  \leq {\bf \mbox{C}}.
\end{equation}
Thus by calculating (or measuring) the channel capacity for SDC, we can draw an inference
about the range in which the entanglement of the shared states (as quantified
by $E_R$) lies. Also one of the bounds, namely, ${\bf \mbox{C}} \leq 1+E_F
$ continues to hold even when we are allowed to vary the a priori probability
of the various signal states (GDC). Thus, our inequality allows one
to impose a readily calculable (from the expression for $E_F$ in Ref.\cite{wot}) limit on the capacity for GDC, without having to optimize GDC over all possible a priori signal probabilities.

    The physical interpretation of the upper bounds becomes clear
from the considerations given in Section.\ref{pff}. During
optimal and complete entanglement purification procedures one generally enhances the classical
capacity much more due to added classical side channels than the inevitable
decrease in capacity brought by the loss of entanglement. The information transferred through these classical side channels 
during purification helps to identify the entangled and disentangled subensembles. It is through this identification that they directly play the
role of boosting the capacity. As after an optimal and complete entanglement purification procedure, the capacity is $1+E_D$, this post-purification
capacity is an upper
bound on the pre-purification capacity. As both the entanglement measures
$E_F$ and $E_R$ are upper bounds on $E_D$, all our upper bounds follow
immediately. It is interesting to note that the upper and the lower bounds
are trivial in complementary regimes. When ${\bf \mbox{C}} \leq 1$, the upper
bounds are trivial, but the lower bound ($E_R \leq {\bf \mbox{C}}$) is
non trivial. On the other hand when ${\bf \mbox{C}} \geq 1$, the upper
bounds are non trivial while the lower bound is trivial.

     We believe that this paper imparts physical significance to the
various measures of entanglement from the viewpoint of forming bounds on
certain kinds of dense coding procedures. This direction of
physical interpretation of the entanglement measures is very different
from the standard interpretations which stem from entanglement
dilution and distillation processes. In a sense, this paper {\em links up
the apparently disconnected notions of entanglement purification
and dense coding}. The considerations of the previous section implies the
following lower bound
on the distillable entanglement: $E_D \geq {\bf \mbox{C}}-1$. This is
important because as yet there is no explicit formula for $E_D$ for a
general mixed entangled state. Thus the readily calculable quantity ${\bf \mbox{C}}-1$ (in the case of SDC) may offer a convinient lower bound on
the distillable entanglement of a given mixed state.

 The interesting question to examine is what apart from entanglement
is involved in determining the dense coding capacity. This may be some measure of distinguishability between
the letter states (hinted by the fact that the average mutual distinguishability
function forms an upper bound on the classical
capacity {\bf C}). An aim of further research
should be to work towards a complete formula for the capacity of dense
coding. Moreover, working with a similar motivation as this paper, one can also try to relate entanglement measures
stemming from purification procedures to the other uses of shared entanglement
such as teleportation \cite{ben93} and secret key distribution \cite{ek91}.

\section{Acknowledgements}
We would like to thank Vladimir Buzek, Daniel Jonathan and Peter Knight for valuable 
discussions. This work is
supported by the Inlaks Foundation, Elsag-Bailey,
Hewlett-Packard, the European TMR networks ERB 4061PL951412 and
ERBFMRXCT96066, the UK Engineering and
Physical Sciences Research Council, the European Science Foundation and the
Leverhume trust.

\newpage

\end{document}